\documentstyle[preprint,aps,pre,eqsecnum,epsfig,amssymb]{revtex}

\begin{document}
\bibliographystyle{prsty}

\draft

\title{\bf Influence of the Soret effect on convection of binary fluids}

\author{St.~Hollinger and M.~L\"ucke}
\address{Institut f\"ur Theoretische Physik, Universit\"at des Saarlandes,
Postfach 151150,\\ D-66041 Saarbr\"ucken, Germany}


\maketitle

\begin{abstract}
Convection in horizontal layers of binary fluids heated from below and in 
particular the influence of the Soret effect on the bifurcation properties
of  extended stationary and traveling patterns that occur for negative
Soret coupling is investigated theoretically.
The fixed  points corresponding to these two convection structures are
determined for realistic boundary conditions with a many mode Galerkin scheme
for temperature  and concentration and an accurate one mode truncation of the
velocity field. This solution procedure yields the stable and unstable
solutions for all stationary and traveling patterns so that complete phase
diagrams for the different convection types in typical binary liquid mixtures
can easily be computed. Also the transition from weakly to strongly nonlinear
states can be analyzed in detail. An investigation of the concentration
current and of the relevance of its constituents shows the way for a
simplification of the mode representation of temperature and concentration
field as well as for an analytically manageable few mode description.
\end{abstract}


\narrowtext

\newcommand{\Nabla}{\mbox{\bf\boldmath $\nabla$}}
\newcommand{\NN}{I\!\!N}
\renewcommand{\vec}[1]{{\bf #1}}
\renewcommand{\u}[1]{\underline{#1}}


\section{Introduction}
Convection in binary miscible fluids like, e. g.,
ethanol--water or $^3$He--$^4$He is a well established and accepted 
system for studying instabilities, bifurcations, complex
spatiotemporal behaviour, and turbulence. This is on the one hand
due to its sufficiently simple experimental realization
under well controllable conditions. On the other hand,
a great advantage for the theoretical
analysis is the solid knowledge of the governing field equations. So,
recently a lot of research activities
\cite{CH93,LL66,PL84,WKPS85,TPC96,AR86,GB86,MS88,EOYSBLKK91,ZM92,WK92,LES96,BLHK89,BLK90,BK90,BLKS95I}
have been directed towards investigating the enormous variety of
pattern forming behaviour in this system. The richness
of spatio temporal phenomena in binary fluid mixtures 
stems from a feed--back loop between the fields of 
velocity, concentration, and temperature. Let us start
with the velocity field: The convective flow is driven
by the buoyancy force field which itself is determined
by variations of the temperature and of the concentration
field. The latter are on the one hand generated via the
thermodiffusive Soret effect by temperature gradients and
on the other hand they are reduced by concentration 
diffusion and by mixing due to the convective flow.
Since these changes influence the buoyancy which drives
the flow the feed back loop is closed. 

In this article we concentrate on the Soret coupling and its influence on
spatially extended convection states of straight parallel rolls that occur
either as a horizontally traveling wave (TW) or in the
form of stationary "overturning" convection (SOC) rolls. 
Among others, we elucidate the Soret induced changes in the combined SOC--TW
bifurcation topology which offers in both types of convection the
possibility of sub-- and supercritically bifurcating branches depending on
the strength of the Soret coupling. Both solution branches develop with
increasing Soret coupling
saddle node bifurcations which give rise to stability changes. Finally,
there exists a merging point of the SOC and the TW branches
for moderate negative Soret couplings. This competition of stationary and
traveling states can only be observed for negative Soret coupling,
where temperature gradients induce adverse concentration gradients
that stabilize the unstable thermal layering. For positive Soret couplings,
there is no oscillatory instability of the basic state. The two interesting
cases
for negative coupling are now {\em weak} and {\em strong} Soret effect. 
For the latter one a bistability of slow and fast TWs was
recently reported \cite{HBL97} which coexist with the likewise stable
basic state of heat conduction. A detailed phase diagram was discussed
\cite{HBL97} describing the Soret dependence of the saddle nodes. For
weak Soret couplings, however, a detailed study of the bifurcation toplogy
was missing. Data from direct numerical simulations are sparse since
in the vicinity of saddle nodes and bifurcation points the intrinsic time
scale of the system is arbitrarily long. Thus, phase diagrams to
elucidate the whole bifurcation topology were incomplete.

We have determined the SOC and TW fixed points of the system by a many mode
Galerkin scheme whose convergence properties do not depend
on the time scale of the system. In particular, this allowed us to study
all unstable branches on which, especially, the
transition from weakly to strongly nonlinear convection takes place.
Furthermore, a detailed explanation of the concentration distribution and
its relation to convection is given. 

Our article is organized as follows: The {\bf second} section presents the
system, the fields needed for its description, their governing equations
with the explanation of the relevant fluid and control
parameters. Finally, it presents a short survey on the typical bifurcation
scenario in the convection of binary liquid mixtures with negative Soret
coupling. The {\bf third} section shows the field truncations, the method
of solving the system and the solutions of our many mode Galerkin scheme
basing on a reasonable approximation in the velocity field.
In the {\bf fourth} section we discuss the influence of the Soret
coupling on the bifurcation topology by means of exemplary bifurcation diagrams
realized in experimentally feasible mixtures and a detailed phase diagram.
Furthermore, evidence for an instability of a TW towards a modulated TW
(MTW) is given. The fluid parameter range for its occurence is elucidated for
ethanol--water as well as for $^3$He--$^4$He--mixtures.
Finally, we extract the relation
between concentration distribution and convective structure and we
investigate the importance of the Soret effect at the boundaries and
its neglibility in the bulk.


\section{System}
A layer of a binary fluid mixture with a mean temperature $\bar T$ and a mean
concentration $\bar C$ is confined between two perfectly heat conducting and
impervious plates separated by a distance $d$ and exposed to a vertical,
homogeneous gravitational acceleration $g$. The lower (upper) plate is
kept at a fixed temperature $\bar{T} +\Delta T/2$ ($\bar{T}-\Delta T/2$).

The fluid parameters are $\rho$ (density of the fluid), 
$\alpha = - \frac{1}{\rho}\frac{\partial\rho}{\partial \bar{T}} $
(thermal expansion coefficient),
$\beta = - \frac{1}{\rho}\frac{\partial\rho}{\partial \bar{C}} $
(solutal expansion coefficient),
$\nu$ (kinematic viscosity),
$\kappa$ (thermal diffusivity),
$k_T$ (thermodiffusion coefficient),
and $D$ (solutal diffusivity).

\subsection{Scaling and balance equations}
We scale lengths by the height $d$ of the layer, times by the vertical
diffusion time $d^2/\kappa$ of the heat and accordingly velocities by
$\kappa/d$. The deviation $T$ of the temperature from its mean $\bar{T}$ is
reduced by $\Delta T$, that of the concentration field by
$\frac{\alpha}{\beta}\Delta T$, and the pressure $p$ by
$\frac{\rho\kappa^2}{d^2}$. Then, the balance equations for mass, momentum,
heat, and concentration \cite{LL66,PL84} read in Oberbeck--Boussinesq
approximation \cite{HLL92}
\begin{mathletters}
\begin{eqnarray}
0 & = & - \Nabla \cdot \vec{u} \label{eq:baleqmass}\\
\partial_t\vec{u} & = & - (\vec{u} \cdot \Nabla) \vec{u}
 - \Nabla\left[p + \left(\frac{d^3}{\kappa^2}g\right) z\right] \nonumber\\
 & & + \sigma \nabla^2\vec{u}
 + R\sigma \left(T+C\right)\vec{e}_z \label{eq:baleqveloc}\\
\partial_tT & = & - \Nabla \cdot \vec{Q}\ =\ 
 - (\vec{u} \cdot \Nabla) T + \nabla^2 T \label{eq:baleqheat}\\
\partial_tC & = & - \Nabla \cdot \vec{J}\ =\
 - (\vec{u} \cdot \Nabla) C + L \nabla^2\left(C -\psi T\right) .
 \label{eq:baleqconc}
\end{eqnarray}
\end{mathletters}
The Dufour effect describing currents of heat driven by concentration
gradients is discarded in (\ref{eq:baleqheat}) since it is relevant
only in binary gas mixtures \cite{HLL92,HL95} or in
liquids near the critical point \cite{LLT83}.

The dimensionless fluid parameters are the Prandtl number
$\sigma=\nu/\kappa$, the Lewis number $L=D/\kappa$, and the separation
ratio $\psi=-\frac{\beta}{\alpha}\frac{k_T}{\bar{T}}$.
The latter characterizes the strength of the
Soret effect. The Rayleigh number
$R=\frac{\alpha g d^3}{\nu \kappa}\Delta T$
serves as control parameter measuring the thermal stress.
 
\subsection{Bounday conditions}
We use experimentally realized boundary conditions for the top and bottom
plates at $z=\pm 1/2$ which are no slip for the velocity field,
$$ \vec{u}(x,y,z=\pm 1/2;t) = 0\ \ \ , $$
perfectly heat conducting for the temperature field,
$$ T(x,y,z=\pm 1/2;t) = \mp 1/2\ \ \ , $$
and impermeable for the concentration field, i.~e.
\begin{equation}
\label{eq:Cbound}
\vec{e}_z\cdot\vec{J} =
 - L\partial_z\left(C-\psi T\right)(x,y,z=\pm 1/2;t) = 0\ \ \ .
\end{equation}
We restrict ourselves to the description of extended roll
like patterns that are homogeneous in one lateral direction
say $y$. So, we investigate two dimensional states of a certain lateral
periodicity length $\lambda=2\pi/k$. In most cases we take $k=\pi$, i.~e.,
$\lambda$ twice the thickness of the fluid layer,
which is close to the critical wavelengths for the negative
Soret couplings investigated here. Furthermore, the stable nonlinear TW and
SOC states that are observed in experiments have typically a wave number
$k=\pi$.

\subsection{Conductive state}
In the motionless basic state, a vertically linear temperature profile,
$T_{\rm cond} = -z$, is observed due to the different top and bottom
temperature. This leads via the Soret effect and the no flux condition
for $\vec{J}$ to a likewise linear concentration profile,
$C_{\rm cond}=-\psi z$. Both together yield the hydrostatic pressure
$$ p_{\rm cond} = p_{\rm 0} - \frac{1}{2} R \sigma (1+\psi) z^2
 - \left(\frac{d^3}{\kappa^2}g\right)z $$
in the quiescent state.

\subsection{Control and order parameters}
The dimensionless temperature difference between the two plates, namely the
Rayleigh number $R$, is used as contol parameter. Mostly we scale it by the
value of the onset of convection in a pure fluid:
$$ r = \frac{R}{R_{\rm c}^{\rm 0}} = \frac{R}{1707.762}\ \ \ . $$

The convective states of the system are characterized by four order
parameters:

(i) The maximum $w_{\rm max}$ of the vertical velocity field.

(ii) The Nusselt number $N = \langle \vec{Q}\cdot\vec{e}_z \rangle_x$
giving the lateral average of the vertical heat current through the system.
In the basic state of heat conduction its value is $1$ and larger than $1$
in all convective states.

(iii) The variance
$M = \sqrt{\langle C^2\rangle_{x,z}/\langle C^2_{\rm cond}\rangle_{x,z}}$
of the concentration field being a measure for the mixing in the system. The
better the fluid is mixed the more the concentration is globally
equilibrated to its mean value $0$ so that $M$ vanishes in optimally
mixing, strongly convecting states.

(iv) The frequency $\omega$ of a traveling wave. Thus, extended TWs
with a wave number $k$ have a phase velocity $v=\omega/k$.
They are stationary states in a reference frame comoving with $v$ relative
to the laboratory system.

\subsection{Typical bifurcation scenario}
For fluid parameters typically realized in mixtures of water and about 10
wt.\% ethanol, an oscillatory, subcritical onset of convection is observed.
It is connected by an unstable TW branch with a saddle node bifurcation
giving rise to stable, strongly nonlinear TW states. At a certain
Rayleigh number, the phase velocity of these waves vanishes and the SOC
branch of stable stationary states can be observed. Along the TW bifurcation
branch which is shown in detail in Fig.~\ref{fig:BifOverview}
the concentration changes its structure from lateral homogeneity and
vertically linear layering in the basic state over plateau--like
distributions in fast TWs to boundary layer dominated, slowly traveling
waves (see the discussion and the figures in Sec.~\ref{sec:IVB}).
The contrast between two adjacent TW rolls is strongly related to
the phase velocity of the TW and it vanishes with this velocity. So, SOCs
do not show such a concentration contrast. In SOCs adjacent rolls are mirror
images of each other and they are separated from another and from the top
and bottom plate only by thin boundary layers. The latter are a typical
phenomenon for convection of weakly diffusing scalars.


\section{Computation of extended states}
\subsection{Modelling the velocity field}
In liquid binary mixtures like ethanol-water momentum
diffuses approximately ten times faster than heat. This means that the
Prandtl number $\sigma$ is of $O(10)$ so that the velocity field may be
adiabatically eliminated. Then, the momentum balance (\ref{eq:baleqveloc}),
say in vertical direction,
reduces to the balance of the diffusive term $\sigma \nabla^2 w$
and the buoyant term $R \sigma \left(T+C\right)$ with the latter
containing no derivatives. Thus, in a stationary flow, either in the
laboratory frame or in a comoving one, the amplitudes of higher lateral
Fourier modes $\hat{w}_n$ of the vertical velocity field 
$w = \vec{u}\cdot\vec{e}_z$ scale at least like $\frac{1}{(nk)^2}$
so that they decrease rapidly and even faster than those of the
temperature field. That is the
reason why higher modes than the critical first lateral
Fourier mode are not necessary for a good description of the
velocity field. One can expect this to hold for all $\sigma\gtrsim 1$.

The next question deals with the role of the lateral mean of
the velocity field, i.~e., of its zeroth lateral Fourier mode. Continuity
implies that only the lateral velocity field $u = \vec{u}\cdot\vec{e}_x$
can contain such a mean flow.
In order to determine the relevance of a $z$--dependent mean flow we
compare its maximum with the two other velocities in the system: The
maximal vertical flow velocity $w_{\rm max}$ and the phase velocity $v$ of
TWs. Just at the onset of convection, the mean flow may be estimated
\cite{LLMN88} to scale with $w_{\rm max}^2$ and to be very
small in comparison with $v$. Furthermore, mean flow and phase velocity have
the same sign. In strongly nonlinear states, the ratio of flow velocity
and mean flow is nearly $10^3$ \cite{BLHK89}, but has changed sign.
Thus, the mean flow has a non--monotonous dependence
on the phase and flow velocity whereas all other properties of TWs
like mixing and heat transport vary monotonically. Hence, the
mean flow cannot contribute systematically to these properties
characterizing TWs sufficiently. This is the
reason, together with the smallness of the mean flow, for ignoring it
altogether.

The $z$--dependence of the critical velocity field is described in an
adequate manner (see, e.~g., \cite{NLK91,LLM91}) by the first even
Chandrasekhar function \cite{CS81} ${\cal C}_1(z)$. Then, the velocity
field of straight rolls with axes oriented in $y$--direction that are
propagating with phase velocity $v$ in $x$--direction is described by
\begin{equation}
\label{eq:uansatz}
\vec{u}(x,z;t) = \frac{w_{\rm max}}{{\cal C}_1(0)}\left(
 \begin{array}{c}
  -\frac{1}{k}\,\sin\,k(x\!\!-\!\!vt)\,{\cal C}_1'(z)\\
  0 \\ \cos\,k(x\!\!-\!\!vt)\,{\cal C}_1(z)
 \end{array}\right)\ \ \ .
\end{equation}
Herein, the phase is chosen so that the maximal vertical flow occurs at
$t=0$ and $x=0$. 

Fig.~\ref{fig:Strofucheck} checks in the first row the applicability of
the ansatz (\ref{eq:uansatz}) by plotting the contribution of modes in the
vertical velocity $w$ which are {\em not\/} represented by (\ref{eq:uansatz})
for two separation ratios $\psi$. The convective amplitude $w_{\rm max}$
was chosen as abscissa to quantify the nonlinearity of the states.
As a measure for the strength of higher contributions to the vertical
velocity the error
\begin{equation}
\label{eq:error}
\left( \frac{  \langle w_{\rm MAC}^2 \rangle_{x,z}
                 -\langle w\, w_{\rm MAC} \rangle_{x,z} }
               {  \langle w_{\rm MAC}^2 \rangle_{x,z}
                 +\langle w\, w_{\rm MAC} \rangle_{x,z} }
\right)^{1/2}
\end{equation}
was computed where $w_{\rm MAC}(x,z)$ denotes the velocity
field calculated from the full field equations
by means of a finite difference MAC scheme \cite{BLKS95I}.
$w(x,z)$ is the one mode approximation (\ref{eq:uansatz}) with the two
velocities $v$ and $w_{\rm max}$ chosen such that the numerically obtained 
velocity $w_{\rm MAC}(x,z)$ is fitted best or, equivalently, the error
(\ref{eq:error}) is minimized.
For both Soret coupling strengths shown in Fig.~\ref{fig:Strofucheck}
the error of the ansatz
(\ref{eq:uansatz}) is smaller than $4$\% but increases as expected
with the convective amplitude.

The bottom row of Fig.~\ref{fig:Strofucheck} shows the maximum of the
mean flow $\langle u(x,z)\rangle_x$. For
both separation ratios the mean flow is smaller than the convective
amplitude by at least three orders of magnitude.

All in all, Fig.~\ref{fig:Strofucheck} justifies the approximations implied by
the ansatz (\ref{eq:uansatz}) for the velocity field. An important consequence
of the fixed spatial structure of the velocity field is that all nonlinear
terms in the balance equations contain now the same amplitude, namely
$w_{\rm max}$, since all nonlinearities are convective ones. This is an
enormous simplification for the theoretical analysis as we will see below.

\subsection{Galerkin expansion for temperature and concentration}
The temperature field $T(x,z;t)$ is appropriately described by
\begin{eqnarray}
\label{eq:Tansatz}
T(x,z;t) = -z & + & 
\sum_{n=0}^{\infty}\sum_{m=1}^{\infty}
  \left[T_{2n}^{2m}(t) e^{-2inkx} + \mbox{c.c.}\right] \sqrt{2} \sin 2m\pi
  z\nonumber\\
 & + & \sum_{n=0}^{\infty}\sum_{m=0}^{\infty}
  \left[T_{2n+1}^{2m+1}(t) e^{-i(2n+1)kx} + \mbox{c.c.}\right] \sqrt{2} \cos (2m
+1)\pi z
  \ \ .
\end{eqnarray}
This representation incorporates the mirror glide symmetry
\begin{equation}
\label{eq:MGS}
\left\{\begin{array}{c}T\\C\end{array}\right\}(x,z;t) = 
-\left\{\begin{array}{c}T\\C\end{array}\right\}(x+\frac{\lambda}{2},-z;t) 
\end{equation}
of TW and SOC states \cite{BLHK89,BLKS95I}. 

The representation of
the concentration field is more subtle because of its boundary condition
(\ref{eq:Cbound}) coupled to the temperature field. The straight forward 
solution is the introduction of the combined field
\begin{equation}
\label{eq:zetadef}
\zeta(x,z;t) = \frac{1}{\psi}C(x,z;t) - T(x,z;t)
\end{equation}
obeying the equation
\begin{equation}
\label{eq:zetabil}
\partial_t\zeta = -\left(\vec{u}\cdot\Nabla\right)\zeta + L\nabla^2\zeta -
\nabla^2T
\end{equation}
and the boundary condition
\begin{equation}
\label{eq:zetabound}
\partial_z\zeta(x,z=\pm1/2;t) = 0\ \ \ \ .
\end{equation}
Note that the $\zeta$ field (\ref{eq:zetadef}) differs by a scaling factor
$1/\psi$ from the field that has mostly been used so far, see,
e.~g., \cite{LL87,LLMN88,HL95}.
An adequate trigonometric expansion is
\begin{eqnarray}
\label{eq:zetaGal}
\zeta(x,z,t) & = &
\sum_{n=0}^{\infty}\sum_{m=0}^{\infty}
  \left[\zeta_{2n}^{2m+1}(t) e^{-2inkx} + \mbox{c.c.}\right] \sqrt{2} \sin (2m+1
)\pi z\nonumber\\
 & + & \sum_{n=0}^{\infty}\sum_{m=0}^{\infty}
  \left[\zeta_{2n+1}^{2m}(t) e^{-i(2n+1)kx} + \mbox{c.c.}\right] \sqrt{2} \cos 2m\pi z
\end{eqnarray}
which also takes the mirror glide symmetry (\ref{eq:MGS}) into account.

The introduction of the combined field $\zeta$ was also motivated by
the wish to fulfill the boundary condition (\ref{eq:Cbound})
or (\ref{eq:zetabound}) for the concentration field exactly.
However, the formulation of the concentration balance in terms of the
$\zeta$ field causes a severe theoretical drawback for small Lewis numbers
$L$ and separation ratios $\psi$ of order $O(1)$, that are typically realized in
liquid mixtures: $\zeta$ and $T$ have the same order
of magnitude according to (\ref{eq:zetadef}). But
in the balance equation (\ref{eq:zetabil}) for $\zeta$ the
diffsuive term enters with weight $L=O(0.01)$ and the temperature with
$O(1)$. This means that for an appropriate solution of
(\ref{eq:zetabil}) for a particular $\zeta$--mode temperature modes are
necessary which are $1/L$ times, i.~e.~about $100$ times, smaller than the
$\zeta$--mode under consideration. Despite the fact that
a relevant contribution of higher
temperature modes was not observed neither in experiments nor in simulations
they are necessary in (\ref{eq:zetabil})
if the $\zeta$--field is introduced and small Lewis numbers are
considered. That is mainly the reason why earlier Galerkin approximations
using the $\zeta$ field and only few temperature modes
did not succeed in describing nonlinear TW convection in binary liquid
mixtures \cite{LLMN88,LLM91,LL87,Y89,L90}. In binary gas mixtures with
typically $L=O(1)$ this problem does not occur \cite{HL95}.

\subsection{Solution of the system of mode equations}
\label{sec:exactresults}
After projecting the balance equations for $T$ and $\zeta$ onto the bases
used in the mode expansions (\ref{eq:Tansatz}) for $T$ and
(\ref{eq:zetaGal}) for $\zeta$ one gets ordinary differential
equations for the mode amplitudes of the temperature and the $\zeta$ field
$$ \vec{X}(t) = \left[ T_{2n}^{2(m+1)}(t) , T_{2n+1}^{2m+1}(t) ,
                    \zeta_{2n}^{2m+1}(t) , \zeta_{2n+1}^{2m}(t)
                    \right]^{\rm T}. $$
The column vector $\vec{X}$ is written here as the transpose of a row vector.

In the case of SOC the amplitudes are
constant and have to be chosen real in order to be compatible with the
velocity ansatz (\ref{eq:uansatz}) for $v\equiv0$. The flow amplitude
$w_{\rm max}$ appears via (\ref{eq:uansatz}) in the $T$ and $\zeta$ field
equations via the convective nonlinearity and in addition as an
inhomogeneous contribution $w\partial_zT_{\rm cond}=-w$ from the conductive
part, $T_{\rm cond}=-z$, of the temperature field (\ref{eq:Tansatz}). Thus,
the mode equations for SOC states take the form
\begin{equation}
\label{eq:SOClinear}
\u{\u{{\cal M}}}_{\rm SOC}(w_{\rm max})\cdot\vec{X}
 = \vec{B}_{\rm SOC}(w_{\rm max})\ \ .
\end{equation}
Here, the matrix $\u{\u{{\cal M}}}_{\rm SOC}(w_{\rm max})$ of mode coupling
coefficients contains $w_{\rm max}$ from the convective nonlinearity.
The momentum balance
in (\ref{eq:baleqveloc}) provides the relation between the Rayleigh number
$R$ and the velocity amplitude $w_{\rm max}$
\begin{equation}
\label{eq:RSOClinear}
R = \frac{w_{\rm max}}{\vec{a}\cdot\vec{X}_{\rm SOC}(w_{\rm max})}
\end{equation}
containing the solution
$\vec{X}_{\rm SOC}=\u{\u{{\cal M}}}_{\rm SOC}^{-1}
 \cdot\vec{B}_{\rm SOC}$ of
(\ref{eq:SOClinear}) depending on $w_{\rm max}$. The vector $\vec{a}$
contains all projection coefficients. The pressure gradient in the momentum
balance of (\ref{eq:baleqveloc}) may be eliminated by taking the curl of
the balance equation. The nonlinearity
of (\ref{eq:baleqveloc}) vanishes in the projection
procedure when using only one velocity mode.
Now, the solution procedure is obvious: Solving the linear system
(\ref{eq:SOClinear}) for a given convective amplitude and inserting
the result into (\ref{eq:RSOClinear}) yields the Rayleigh number of that
SOC state with the convective amplitude $w_{\rm max}$. Thus, the
bifurcation diagram $R(w_{\rm max})$ or $w_{\rm max}(R)$ may be calculated.

Finding the TW solution is slightly more difficult since the modes
are time dependent:
$$ \vec{X}_{\rm TW}(t) = \left[ \widehat{T}_{2n}^{2(m+1)}e^{i2n\omega t} ,
                     \widehat{T}_{2n+1}^{2m+1}e^{i(2n+1)\omega t} ,
                     \widehat{\zeta}_{2n}^{2m+1}e^{i2n\omega t} , 
                     \widehat{\zeta}_{2n+1}^{2m}e^{i(2n+1)\omega t}
              \right]^{\rm T}\ \ . $$
The form of the time dependence is determined by the fact that the TWs are
stationary in a frame comoving with $v=\omega/k$. Therefore, the angular
frequency of the $n$th lateral Fourier mode (lower index of the mode
amplitudes) is $n\omega$ with $\omega$ being the basic frequency of
the TW. With the complex vector of TW mode amplitudes
$$ \widehat{\vec{X}}_{\rm TW} = \left[ \widehat{T}_{2n}^{2(m+1)} ,
                              \widehat{T}_{2n+1}^{2m+1} ,
                              \widehat{\zeta}_{2n}^{2m+1} ,
                              \widehat{\zeta}_{2n+1}^{2m}
                       \right]^{\rm T} $$
the system of the projected balance equations reduces once more to a
linear algebraic system
\begin{equation}
\label{eq:TWlinear}
\u{\u{{\cal M}}}_{\rm TW}(\omega,w_{\rm max})\cdot\widehat{\vec{X}}_{\rm TW}
= \vec{B}_{\rm TW}(w_{\rm max})
\end{equation}
where, however, the matrix $\u{\u{{\cal M}}}_{\rm TW}$ of mode coupling
coefficients is complex.
Another complex equation (or two real ones) generated by the momentum
balance relates the two real groups $\omega$ and $w_{\rm max}$ with the
Rayleigh number $R$ and the separation ratio $\psi$
\begin{equation}
\label{eq:RTWlinear}
\left(
\begin{array}{cc}
 \vec{a}_{11}\!\!\cdot\!\!\widehat{\vec{X}}_{\rm TW}
  & \vec{a}_{12}\!\!\cdot\!\!\widehat{\vec{X}}_{\rm TW}\\
 \vec{a}_{21}\!\!\cdot\!\!\widehat{\vec{X}}_{\rm TW}
  & \vec{a}_{22}\!\!\cdot\!\!\widehat{\vec{X}}_{\rm TW}
\end{array}
\right)
\left( \begin{array}{c} 1/R \\ \psi \end{array} \right) = 
\left( \begin{array}{c}
        \vec{b}_{1}\!\!\cdot\!\!\widehat{\vec{X}}_{\rm TW}\\
        \vec{b}_{2}\!\!\cdot\!\!\widehat{\vec{X}}_{\rm TW}
       \end{array}
\right)
\end{equation}
with $\vec{a}_{ik}$, $\vec{b}_{i}$ being vectors of projection
coefficients. 

A possibility to solve the system is to solve (\ref{eq:TWlinear})
for a given combination ($\omega,w_{\rm max}$) and to use the resulting
$\widehat{\vec{X}}_{\rm TW}(\omega,w_{\rm max})$ for solving (\ref{eq:RTWlinear}) with
resepect to ($1/R,\psi$). This means that $R$ and $\psi$ are uniquely
determined for a given combination ($\omega^2,w_{\rm max}^2$).
The relevant parameters are here the squares $\omega^2$ and $w_{\rm max}^2$
since left-- and right--traveling waves are symmetry degenerated and the
convective amplitude $w_{\rm max}$ was chosen to be positive by
(\ref{eq:uansatz}).

This result is illustrated in Fig.~\ref{fig:omwmax}: The TW
states fill the $\omega^2$--$w_{\rm max}^2$--plane. The lines are connecting
states along a TW bifurcation branch for a particular separation ratio.
The control parameter $R$ varies along a particular line
between $R_{\rm osc}$ at the Hopf bifurcation ($w_{\rm max}=0$)
and $R^*$ at the SOC--TW--transition ($\omega^2=0$) in a way that can be
non--monotonous. The value of $\omega^2$ for
$w_{\rm max}^2=0$ is $\omega_H^2$, i.~e., the square of the Hopf frequency
$\omega_H$. The dotted line represents the position of the saddle nodes in
the diagrams $\omega(R)$ or $w_{\rm max}(R)$. It vanishes in the vicinity
of $\psi=-0.01$ since for weaker Soret couplings only an unstable lower TW
bifurcation branch exists without any saddle node. These
topological features will be discussed later in this article.

\subsection{Comparison with finite difference numerical results}
In Fig.~\ref{fig:BifVgl} we give a quantitative comparison of bifurcation
diagrams computed by the above presented numerical Galerkin method
and those obtained by a finite difference MAC scheme with a
full representation of the fields. In our Galerkin scheme, we used one
velocity mode and up to $19$ temperature and $\zeta$ field modes in each
direction. So, we computed $761$ real mode amplitudes in the case of TW
convection.

For both SOC and TW states the
pairs of diagrams coincide. The most evident discrepancy can be observed in
the end point $r^*$ of traveling wave convection in
Fig.~\ref{fig:BifVgl}(d): The MAC results with a discretization of
$dx=dz=0.05$ (solid circles) predict $r^*\simeq1.65$ \cite{BLKS95I}, the
Galerkin scheme (dashed line) $r^*\simeq1.495$. In order to elucidate this
deviation we performed a finite difference calculation with $dx=dz=0.025$
(open lozenges) and observed $r^*\simeq1.45$, i.~e., close to the value of
the Galerkin scheme which used modes up to a wave number $19\pi$ in both
directions. The variation of $r^*$ with the spatial resolution of the MAC
scheme is caused by concentration boundary layers at the plates:
Galerkin method and the finer resolving MAC scheme show the same thickness
which is smaller than that
predicted by the worse resolving method $dx=dz=0.05$. Now, the
SOC--TW--transition may be interpreted as a boundary layer instability
\cite{BPS89} that occurs if the SOC boundary layer thickness
$\propto(L/w_{\rm max})^{1/3}$ exceeds a certain value when reducing $r$
or reducing the convective amplitude. This critical
thickness is reached for higher amplitudes when using a numerical method
that produces larger boundary layers. In so far, $r^*\simeq1.45$ is
a more adequate value for $L=0.01$, $\sigma=10$, and
$\psi=-0.25$ than $r^*\simeq1.65$ \cite{BLKS95I}.

The deviations in the bifurcation diagrams for the variance $M$ of the
concentration in Fig.~\ref{fig:BifVgl}(b,e) are mainly due to the
shift in the frequency bifurcation branches. This means that the
dependence of the concentration distribution on the frequency is reproduced
well.

The convective heat transport $N-1$ [Fig.~\ref{fig:BifVgl}(c,f)] in our
Galerkin approximation is carried by only one velocity mode so that the
actual values may be expected to be larger the higher the forcing,
i~.e., the Rayleigh number $r$. This typical behaviour can be observed also
for the pure fluid $\psi=0$. There, a discrepancy between full MAC results
and our Galerkin approximation of about $6$\% at $r=2$ is seen.

\section{Results}
\subsection{Soret coupling and bifurcation topology}
The dependence of the bifurcation topology on the strength of the Soret 
coupling, say in the range $-0.25<\psi<0$, was
not yet discussed in the literature
\cite{BLK90,BLKS95I,BPS89} in detail: On the one hand,
the directly integrating numerical methods
\cite{BLK90,BLKS95I} require large amounts
of CPU time due to critical slowing down near the saddle node positions 
$r_{\rm SOC}^s$ and $r_{\rm TW}^s$. On the other hand, the Soret effect was
only implemented incompletely in a theoretical approach\cite{BPS89}.
The same applies to the transition
point $r^*$ of TW to stationary convection. The Galerkin method
presented in the last section does not suffer from these disadvantages and
it yields also the whole unstable branches directly
without any numerical tricks. Since it computes only fixed points
the intrinsic time scale of the system does {\em not\/} enter the
problem. However, an additional
stability analysis of the computed states is necessary.

\subsubsection{Strong negative $\psi$}
The investigation of the variation of the bifurcation scenario with the
separation ratio $\psi$ has shown for {\em strong\/} negative couplings
a very interesting feature \cite{HBL97}: the development of a lower stable
TW branch out of the bump seen in Fig.~\ref{fig:BifOverview} in the unstable
branch for $\psi=-0.25$. For $\psi\lesssim-0.4$ two stable, convective
TW solutions exist opening up the possibility of the coexistence
of two different traveling states in {\em one\/} container. Furthermore,
all these TWs were found to display universal scaling properties
\cite{HBL97}: (i) The mixing $M$, i.~e., the concentration variance varies
linearly with the TW frequency. (ii) The latter itself is uniquely fixed by
the "distance", $r(w_{\rm max},\psi)-r(w_{\rm max},\psi=0)$, of the TW
state's location ($r,w_{\rm max}$) on the TW bifurcation branch from the
pure fluid ($\psi=0$) convection coordinates in the $r$--$w_{\rm max}$
bifurcation diagram. Thus, bifurcation properties, spatial structure of the
concentration distribution, and TW dynamics show for strong Soret coupling
a peculiar scaling behaviour. For a discussion of the
characteristic changes in the bifurcation topology at {\em strong\/}
Soret coupling see Ref.~\cite{HBL97}.

\subsubsection{Weak negative $\psi$}
On the side of {\em weak\/} negative Soret couplings,
i.~e., $\psi\rightarrow 0$, the motion of the saddle nodes
$r_{\rm SOC}^{\rm s}$ and $r_{\rm TW}^{\rm s}$ and of the SOC--TW merging
point $r^*$ in the $r$--$\psi$ plane was not elucidated except for the
vanishing of stable TW convection for $\psi>-0.01$
in mixtures with ethanol--water parameters $L=0.01$ and $\sigma=10$
\cite{BLKS95I}. To fill this gap we discuss in
Figs.~\ref{fig:BifTop}--\ref{fig:PhaseGanz} the bifurcation properties of
TWs and SOCs in the range $-0.25<\psi<0$. In Fig.~\ref{fig:BifTop} we show
TW and SOC bifurcation diagrams of $w_{\rm max}^2$ {\it vs\/} $r$ for
several $\psi$ as indicated.
In the case of the strongest coupling $\psi=-0.05$
[Fig.~\ref{fig:BifTop}(a)]
the same situation as for $\psi=-0.25$ (Fig.~\ref{fig:BifOverview}) is observed
except for the fact that here $r^*<r_{\rm osc}$ so that a SOC
state is observed when heating above threshold. The stationary bifurcation
threshold, $r_{\rm stat}$, is negative as it is the case for all
$\psi<\frac{-L}{1+L}=-\frac{1}{101}$ for $L=0.01$ \cite{LLT83,HL95}: Thus, the
SOC branch is disconnected with the ground state $w_{\rm max}^2\equiv 0$
at the positive $r$--axis. The shape of the SOC branch does not
change when reducing the strength of the Soret effect because the
tricritical separation ratio for SOCs, $\psi_{\rm SOC}^t$,
scales with $-L^3$ \cite{SZ93} and is effectively $0$ for
small Lewis numbers. The most evident effect is the motion  of the
SOC--TW--transition point $r^*$ along the SOC branch towards the SOC
saddle node. For $\psi=-0.02$ [Fig.~\ref{fig:BifTop}(b)], the TW and SOC
saddle nodes coincide; for $\psi=-0.01$ [Fig.~\ref{fig:BifTop}(c)] the TW
branch merges with the SOC branch at the SOC
saddle node. The transition from SOC to a pair of symmetry degenerated
TWs at $r^*$ which is a pitch fork
bifurcation of the TW frequency is still
backwards in the sense that the frequency bifurcation is subcritical.
Different to $\psi=-0.02$ we can observe for $\psi=-0.01$ TWs also for
control parameters smaller than those of all SOCs, i.~e., the
extended state with the smallest Rayleigh number is now a TW and no longer a
SOC. This is different from the behaviour for
$\psi<-0.02$ where SOCs exist also at Rayleigh numbers not allowing TWs.

In Fig.~\ref{fig:BifTop}(d) with $\psi=-0.0085$ the TW
branch merges with the lower, unstable SOC branch
which then becomes stable beyond its
saddle node. But the TW saddle is still found at smaller amplitudes and
Rayleigh numbers than that of the SOCs. The consequences are as follows: The
lower TW branch locates unstable TWs which become stable at the
saddle. However, they change stability once more since stable TWs
may not merge with the unstable SOC branch. The only possibility is the
existence of an additional bifurcation point --- here to a modulated TW
(MTW) --- on the upper TW branch between  $r_{\rm TW}^{\rm s}$ and $r^*$.
This scenario is investigated in more detail in the next subsection
\ref{sec:IVA3}.

Here, we continue the discussion of the changes in the bifurcation topology
of ethanol--water mixtures for $\psi\rightarrow 0$. At $\psi=-0.007$ in
Fig.~\ref{fig:BifTop}(e) the TW saddle has vanished and the whole TW branch
is unstable. The next significant
change in the topology occurs at the tricritical separation ratio
$\psi_{\rm TW}^t\simeq -5\!\cdot\!10^{-5}$ \cite{SZ93}. Therefore,
at $\psi=-4\!\cdot\!10^{-5}$ we observe in Fig.~\ref{fig:BifTop}(f)
a supercritically bifurcating TW branch. It is stable just at the onset
and then becomes unstable with respect to MTWs. At the codimension--$2$--point
$\psi_{\rm CT}=-3.526\!\cdot\!10^{-5}$ \cite{SZ93} the TW branch vanishes
completely. For separation ratios $\psi$ between $\psi_{\rm TW}^t$ and
$\psi_{\rm CT}$ there is also a necessity of a change in stability along the
supercritically bifurcating TW branch ending on the unstable SOC branch.

\subsubsection{Modulated TWs}
\label{sec:IVA3}
We have checked the scenario for the appearence of
MTWs more explicitly for a parameter combination
which is realized in $^3$He--$^4$He--mixtures rather than in ethanol--water,
namely $L=0.03$, $\sigma=1$, and $\psi=-0.055$. In $^3$He--$^4$He this
scenario occurs in a four times broader range of the control parameter
($r\in[1.0612,1.0614]$, see Fig.~\ref{fig:MTW}(a)) than it is realized in
ethanol--water mixtures ($r\in[1.01375,1.01380]$, see
Fig.~\ref{fig:BifTop}(d)). 

Sufficiently above the
oscillatory threshold $r_{\rm osc}=1.08827$ ($\omega_H=3.8467$,
$k_c=3.1152$) stable SOCs (filled triangles) can be realized according
to Fig.~\ref{fig:MTW}(a). When reducing
the heating rate $r$ we approach the SOC--saddle at about
$r_{\rm SOC}^s\simeq1.06139$. Below that, the system falls down on the
branch of TWs which is oscillatory unstable (open squares) at
$r_{\rm SOC}^s$. The upper stable TW branch (filled circles) turns unstable
at $r_{\rm MTW}\simeq1.06125$ shortly above
the saddle $r_{\rm TW}^s$. The oscillatory unstable TWs (open squares)
get modulated by a frequency at least ten times smaller than the basic
frequency of the TW itself --- compare the frequencies in
Fig.~\ref{fig:MTW}(b) and the imaginary parts of the relevant eigenvalues in
Fig.~\ref{fig:MTW}(c). The instability of the TWs on their upper branch,
i.~e., the transition at $r_{\rm MTW}$ from filled circles
to open squares, gives birth to
MTWs with slowly breathing amplitude. The type
of the instability of the TWs changes on the upper TW branch at
$r\simeq1.06141$ from oscillatory to stationary (open circles) in order to
merge appropriately with the stationary unstable lower SOC branch (open
triangles).

This scenario has also
been discussed in a minimal model by Knobloch and Moore \cite{KM90}. They,
however, investigated a system with stress free and, more importantly,
permeable boundaries without adequately resolving the boundary layers.
We give here evidence for MTWs in binary mixtures with realistic
boundary conditions realized in the experiments. Note, however,
the experimental investigation requires a control of the
temperature difference of about $10^{-4}$.

\subsubsection{Phase diagram}
Our results are summarized in a phase diagram in
Fig.~\ref{fig:PhaseGanz} where the $\psi$--dependence of $r^*$ (open
lozenges), SOC saddle (open triangles up), and TW saddle (filled circles)
are displayed together with the linear stability thresholds of the basic
state (stationary: open triangles down, oscillatory: filled squares).
Stable TW states on the upper TW bifurcation branch are located in the
shaded region of Fig.~\ref{fig:PhaseGanz}.
The inset in the upper right corner covers separation ratios
between $\psi=-0.05$ and $\psi=-0.005$. In this inset we have
scaled $r-1$ by $\sqrt{-\psi}$ in order to map the SOC saddle node position
$r_{\rm SOC}^s$ approximately onto a constant since its $\psi$--dependence
may be fitted very well by
\begin{equation}
\label{eq:rSSOC}
r_{\rm SOC}^s \simeq 1 + 1.636 \sqrt{-L\psi}\ \ \ \ .
\end{equation}
However, in view of the fact that the exponent in (\ref{eq:rSSOC}) is
not exactly $1/2$ it is not surprising that $r_{\rm SOC}^s$ is not exactly
constant in the inset of Fig.~\ref{fig:PhaseGanz}. The range of existence
of the MTWs is too small to be visible in Fig.~\ref{fig:PhaseGanz}.

The changes of the bifurcation topology induced by increasing the strength
of the negative Soret coupling beyond values of $-0.25$ including
a detailed phase diagram for $\psi\in [-0.7,-0.2]$ were discussed in
Ref.~\cite{HBL97}: There, two stable, nonlinear TW branches have been
discovered for $\psi\lesssim -0.4$ in mixtures with $L=O(0.01)$ and
$\sigma=O(10)$.

\subsection{Concentration distribution and streamfunction}
\label{sec:IVB}
Fig.~\ref{fig:GeschwBif} shows a combined bifurcation diagram of convective
amplitude $w_{\rm max}$ and phase velocity $v$ {\em vs.\/}~reduced Rayleigh
number $r$. The TW states have been computed with our Galerkin method using
the one-mode velocity field approximation (\ref{eq:uansatz}).
Four states are labelled by A to
D in agreement with Fig.~\ref{fig:BifOverview}. 

We want to discuss here the changes in the concentration distribution when
moving along the TW bifurcation branch and their relation to the structural
changes of the streamfunction 
\begin{equation}
\label{eq:strofumit}
\tilde{\phi}(x,z) = \frac{w_{\rm max}}{k\,{\cal C}_1(0)}
 \,\sin\,kx\,{\cal C}_1(z) + v\,z
\end{equation}
for the velocity field (\ref{eq:uansatz}) in the frame that is comoving
with the TW phase velocity $v$. Note that in this frame the velocity field
is stationary so that passive particles would move along the streamlines of
(\ref{eq:strofumit}).

The first occurrence of local extrema in $\tilde{\phi}(x,z)$
(\ref{eq:strofumit}) gives rise to the first appearence of areas of closed
streamlines. That happens for $k=\pi$ at the velocity ratio
$\chi = \frac{w_{\rm max}}{v} = 0.9807$. Thus, for $\chi\lesssim 1$, i.~e.,
in all TW states between the Hopf bifurcation threshold $r_{\rm osc}$ and
state A the streamlines are all open.
Increasing the convective amplitude
$w_{\rm max}$ yields decreasing phase velocity $v$ and 
increasing values of $\chi$ so that areas of closed streamlines appear that
grow on cost of those occupied by open streamlines. For $v=0$, i.~e., at the
point D only closed streamlines are observed.

Without feedback into buoyancy concentration is a passive scalar transported 
by means of convection and
diffusion (for the discussion of the Soret coupling in the bulk equations see
Sec.~\ref{sec:Soretcoupl}). Then
one can apply the results of passive scalar theory, e.~g.,
\cite{RY83,S87}. It explains that within closed streamlines a weakly
diffusing scalar is homogenized. This behaviour is elucidated in
Fig.~\ref{fig:Profile} where we show lateral and vertical concentration
profiles for the states A -- D labelled in Fig.~\ref{fig:GeschwBif}. In the
states B, C, and D where $\chi>1$ and therefore areas of open streamlines
exist we see a plateau characteristic of both profiles. However, the state
A with only open streamlines shows no such plateaus. 

Note that the lateral concentration wave 
profiles between the Hopf bifurcation and the state A are basically
harmonic and that their amplitudes
increase with increasing convective amplitude. This amplitude growth of the
harmonic concentration wave occurs as long as there are not yet closed
streamlines, i.~e., up to a limiting point in the vicinity of state A. The
amplitude of the wave profile of the state B in Fig.~\ref{fig:Profile} is
larger than that of state A as the lateral profiles are not taken at the
center of the closed streamlines but in the center of the convection cell.
Beyond this maximum amplitude the concentration wave crests are "cut off"
and a plateau develops with the appearence of closed streamlines. The
plateau extension, i~.e., the region of constant concentration within the
areas of closed streamlines broadens with increasing $\chi$ while
simultaneously the plateau value of the concentration wave reduces.
Furthermore, a small concentration peak at the leading front of the right
traveling wave develops. It is caused by the advective injection of the
concentration at the plates into the areas of closed streamlines.
The lateral profile of state A has only a very weak, not visible
asymmetry with respect to reflection at $x=\pm1/2$ whereas the asymmetry in
state B is obvious and gets more and more pronounced in C and D. 
The adevctive injection of concentration into the regions of closed
streamlines takes --- with diffusion being small compared to advection ---
the form of an inwards spiralling concentration jet that can be seen in
Fig.3 of Ref.~\cite{BLKS95I}.

This strong relation between streamlines and concentration distribution
is also demonstrated in Fig.~\ref{fig:CvonPhi} where we have plotted the
concentration field $C$ {\em vs.\/}~the streamfunction $\tilde{\phi}$
in the frame comoving with the TW state traveling to the right. Just at the
Hopf bifurcation threshold the streamfunction in the comoving frame is
$\tilde{\phi} = v z = \omega_H z/k$ reflecting a vanishing convective velocity
field in the laboratory frame. Via the conductive concentration profile 
$C_{\rm cond}=-\psi z$ we get the relation
$C_{\rm cond}(\tilde{\phi})=-\frac{k\psi}{\omega_H} \tilde{\phi}
 \simeq 0.070 \tilde{\phi}$
(dotted lines in Fig.~\ref{fig:CvonPhi})
between concentration and streamfunction with $\tilde{\phi}$ varying between
$-v/2$ and $v/2$, i.~e., between $-1.78$ and $1.78$.
In state A, the conductive concentration distribution is slightly
deformed and the streamfunction varies between $-1.4$ and $1.4$ which are
the values at the top and bottom plates. At these boundaries the concentration
is in contrast to the streamfunction
not constant so that a vertical shape of $C(\tilde{\phi})$ is
observed there. In state B there are already small
areas of closed streamlines with equilibrated
concentration. In these regions $C(\tilde{\phi})$ is a constant.
Moving along the TW bifurcation branch the 
areas of these regions keep on increasing via state C to state D which is a SOC
state. In this state we observe a remarkable concentration variation around
its mean value $0$ only in a small streamfunction interval
around $0$, the separatrix in this state. This is a strong boundary layer
phenomenon caused by the smallness of Lewis number $L=O(0.01)$
in comparison with the convective amplitude $w_{\rm max}=O(10)$.

\subsection{Soret coupling and concentration current}
\label{sec:Soretcoupl}
\subsubsection{Comoving frame of reference}
The impact of the Soret coupling on the concentration current in a
TW is best understood by studying the current in a frame
comoving with the TW's phase velocity $v$. In this
frame, the velocity field 
$\tilde{\vec{u}}(x,z)=(-\partial_z \tilde{\phi},0,\partial_x \tilde{\phi})$
is well approximated by the ansatz (\ref {eq:strofumit})
according to ansatz (\ref{eq:uansatz}) and the discussion of
Fig.~\ref{fig:Strofucheck}. The corresponding streamlines can be seen in
Fig.~\ref{fig:Stroeme}(a) for a TW state with $v=4/\pi$, i.~e., $\omega=4$.

In the comoving frame the concentration current $\tilde{\vec{J}}$
is given by
\begin{equation}
\label{eq:conccurrent}
\tilde{\vec{J}} = \tilde{\vec{u}}C - L\Nabla(C-\psi T)\ \ .
\end{equation}
In a relaxed TW state the relation
$$ \partial_tC = 0 = - \Nabla\cdot\tilde{\vec{J}}$$
holds so that $\tilde{\vec{J}}$ is divergence free.
It is shown as a vector field plot in Fig.~\ref{fig:Stroeme}(a) in the left
column together with the streamlines of the velocity field
$\tilde{\vec{u}}$ for a TW propagating to the right.
The structure of the current can be
explained as follows: In the closed streamlines of the right rolls of
Fig.~\ref{fig:Stroeme} concentration is homogeneized to a level close to
that of the upper plate. This is due to the vicinity of the right roll to
this plate where the higher concentration ($C>0$) is observed for negative
Soret couplings and heating from below. The
left area of closed streamlines is located next to the lower plate and
contains therefore the lower concentration ($C<0$). The two different
signs of concentration --- the mean concentration is normalized to zero ---
are the reason for the concentration current
$\tilde{\vec{J}}$ to rotate
clockwise in both clockwise and counterclockwise rotating rolls. In the
vicinity of the center open streamline that meanders between the rolls,
a line of vanshing concentration
current exists. It also vanishes near the center of the rolls where the
velocity $\tilde{\vec{u}}$ vanishes.

At the plates the concentration current is purely
lateral since due to the impermeable boundary condition
$\tilde{\vec{J}}\cdot\vec{e}_z=0$. It is
mainly given by $-vC$ being negative (positive) at the top (bottom) plate
with $C>0$ ($C<0$). The contribution from lateral
concentration gradients along the plates to the
concentration current (\ref{eq:conccurrent}) is mutiplied by the Lewis
number $L=O(0.01)$ and can therefrore be neglected there. The temperature, and
consequently the Soret effect, does not contribute at all to
$\tilde{\vec{J}}\cdot\vec{e}_x$ at the plates since the temperature
is fixed there.

\subsubsection{Approximate Soret induced current}
Next, we discuss the influence of the temperature field on the
concentration current (\ref{eq:conccurrent}) in the bulk.
We do this by studying the
left column of Fig.~\ref{fig:Stroeme}(b). There, $L\psi\Nabla T$ is plotted
by arrows whose lengths are magnified by a factor $30$ relative to that of
$\tilde{\vec{J}}$ in Fig.~\ref{fig:Stroeme}(a).
For negative Soret coupling $L\psi\Nabla T$ is parallel
to the diffusive heat current $-\Nabla T$ pointing upwards in
the system heated from below. The important thing is the
existence of this current and its mean upwards direction and not
the small modulations. We ignore these small lateral modulations by replacing
$L\psi\Nabla T$ by an adequate mean. For this mean we choose the mean of
$L\psi\Nabla T$ at the plates, namely
\begin{equation}
\label{eq:meanSoret}
\langle L\psi\Nabla T\rangle_{x,z=\pm1/2} = -L\psi N \vec{e}_z\ \ ,
\end{equation}
in order to guarantee the impermeability of the plates in the lateral mean.
A global averaging of $L\psi\Nabla T$ would lead to $-L\psi\vec{e}_z$
differing from our choice only by a factor $N=O(1)$ and violating the
impermeability by the same amount.

The replacement of $L\psi\Nabla T$ by the mean (\ref{eq:meanSoret}) leads
to the modified concentration current
\begin{equation}
\label{eq:modcurrent}
\tilde{\vec{J}}_{\rm modified} = \tilde{\vec{u}}C - L\Nabla C - L\psi
 N\vec{e}_z\ \ 
\end{equation}
shown in the right column of Fig.~\ref{fig:Stroeme}(a). Therefore, in the 
modified concentration balance
\begin{equation}
0 = \partial_tC = -\Nabla\cdot\tilde{\vec{J}}_{\rm modified}
  = -\Nabla\cdot\left[\tilde{\vec{u}}C - L\Nabla C \right]
\end{equation}
the tempreature field has disappeared. It 
occurs only in the boundary condition
$$ 0 = \vec{e}_z\cdot\tilde{\vec{J}}_{\rm modified} = 
     -L\partial_zC - L\psi N\ \ \ \ \text{at }z=\pm1/2\ \ .$$
This allows the concentration field to be described by
\begin{eqnarray}
C(x,z;t) = - \psi N z & + &
\sum_{n=0}^{\infty}\sum_{m=0}^{\infty}
  \left[C_{2n}^{2m+1}(t) e^{-2inkx} + \mbox{c.c.}\right] \sqrt{2} \sin (2m+1
)\pi z\nonumber\\
 & + & \sum_{n=0}^{\infty}\sum_{m=0}^{\infty}
  \left[C_{2n+1}^{2m}(t) e^{-i(2n+1)kx} + 
  \mbox{c.c.}\right] \sqrt{2} \cos 2m\pi z\ \ .\nonumber
\end{eqnarray}
Since now the actual concentration balance is solved and not the balance for
the $\zeta$--field the above discussed problems with the relevance of
temperature modes that are $1/L$ times smaller than the $\zeta$--field modes
do no longer occur. Consequently, the temperature field can now be
represented by the simplest approximation
\begin{eqnarray}
\label{eq:Tansatzred}
T(x,z,t) = -z & + &
T_0^2(t)\sqrt{2}\sin 2\pi z\nonumber\\
 & + & \left[T_1^1(t) e^{-ikx} + 
  \mbox{c.c.}\right] \sqrt{2} \cos \pi z
\end{eqnarray}
just like in the classical Lorenz model \cite{L63}.
The advantage of this procedure is that only for the concentration field a
many mode Galerkin representation is necessary and the temperature can
be modelled by those modes which are observed in the fields and not
additionally by those becoming necessary out of numerical reasons.

Bifurcation diagrams of the frequency computed by this
approximate method (solid lines)
are displayed in Fig.~\ref{fig:RandBulk} and compared with the
"exact" results (symbols) as described in Sec.~\ref{sec:exactresults}
over a wide range of Soret
couplings. We have chosen bifurcation diagrams for the frequency since it
is most sensitive to an insufficient mode truncation. For separation
ratios $\psi>-0.15$ no differences are observed whereas for stronger
couplings ($\psi=-0.25$) at high heating rates deviations become visible.
The reason for this is the neglect of temperature 
modes higher than those incorporated in (\ref{eq:Tansatzred}). At
high heating rates they become relevant: The maximum
of the second lateral temperature mode has reached
for $r=1.6$ about $6$\% of the size of the first
one ($3$\% for the third relative to the first) (see also
\cite[Fig.5b]{BLKS95I}). In so far, the simplest ansatz for the temperature
field (\ref{eq:Tansatzred})
cuts off at least $10$\% of the spectrum leading to comparable errors in
the bifurcation diagram, especially in the vicinity of the
SOC--TW--transition.

\section{Conclusion}
We have given a detailed analysis of the influence of the Soret effect on
thermal convection in binary liquid mixtures. As a tool we used a many mode
Galerkin method combined with an approximation in the velocity field: It is
truncated by a single mode which holds for convection in fluids with
Prandtl numbers $\sigma \gtrsim 1$. The mean flow in traveling waves (TWs) was
recognized as unimportant since it is very small and contributes only
unsystematically to the main TW properties. This truncation of the velocity
lead to a very simple and efficient solution procedure for the nonlinear
TW and SOC fixed points by only solving linear equations. This opened up the
possibility for a detailed elucidation of
the changes in the combined bifurcation topology of stationary and
traveling states, especially the existence range of TWs in the
control and fluid parameter plane. Together with the investigation of
strong Soret couplings \cite{HBL97} interesting fields for experiments
are opened: the bistability of slow and fast traveling waves and the
occurrence of modulated traveling waves with slowly breathing
amplitude. The classification of the states is facilitated by a detailed
phase diagram. As an additional insight it was found that the order paramters
TW frequency and TW convective velocity determine the control parameters
Rayleigh number and separation ratio uniquely.

Eliminating the intrinsic time scale from the computation of the fixed points
allowed besides the determination of bifurcation points and saddle nodes
the investigation of unstable states. In this regime the transition from
weakly to strongly nonlinear TWs is observed and may be understood
in the framework of the intimate relation between concentration distribution
and the structure of the flow and the changes in this relation along the
bifurcation branch. 

As a further result the reason for the failure of earlier Galerkin
approximations for the convection in binary liquid mixtures was revealed:
In order to {\em exactly\/} fulfill the concentration boundary
condition which is coupled to the temperature field earlier approaches used
a combination of concentration and temperature field without resolving the
temperature field adequately. When using this combined field for small
Lewis numbers $L \ll 1$ a resolution of the temperature is required that
goes beyond the dominating modes seen in simulations and experiments.
Although the contribution of higher modes to the temperature field are small
they are essential in the balance equations formulated with the combined
field. A way out of this artificial theoretical dilemma is obtained by
investigating the concentration current: It is mainly influenced by the lateral
average of the temperature gradient in the system. This allows to ignore the
Soret effect in the bulk equation of the concentration balance (bulk Soret
effect) and to truncate the concentration field with an adequate boundary
condition (boundary Soret effect) directly. Then, the foundations of a more
compact description and solution for the convection in binary liquid mixtures
are laid \cite{HLM97}.

\acknowledgements
This work was supported by the Deutsche Forschungsgemeinschaft. Stimulating
discussions with W.~Barten, P.~B\"uchel, and H.~W.~M\"uller are gratefully
acknowledged.

\begin{figure}
\caption{Bifurcation diagrams of convection flow intensity (a)
         and frequency (b) in TW (solid lines) and SOC
         (dashed lines) solutions in a binary mixture with $\psi=-0.25$,
         $L=0.01$ and $\sigma=10$. States A--D are identified for later
         discussion related to 
         Figs.~\ref{fig:GeschwBif}-\ref{fig:CvonPhi}.}
\label{fig:BifOverview}
\end{figure}

\begin{figure}
\caption{Quality of the one--mode velocity field approximation
         (\ref{eq:uansatz}) for TWs with $L=0.01$, $\sigma=10$ and $\psi=-0.25$
         (left column) and $\psi=-0.6$ (right column) as a function of growing
         flow amplitude $w_{\rm max}$ along the TW bifurcation branches. The
         top row displays the error
         in the vertical velocity field according to definition (\ref{eq:error})
         and the bottom row the maximal amplitude of the lateral mean flow.}
\label{fig:Strofucheck}
\end{figure}

\begin{figure}
\caption{TW states in the $w_{\rm max}^2$--$\omega^2$--plane
         for $L=0.01$ and $\sigma=10$. The separation ratio $\psi$ is constant
         on each solid line. It varies logarithmically
         form line to line and has the value
         shown at the left ordinate. The dotted line gives the positions of
         the TW saddle nodes.}
\label{fig:omwmax}
\end{figure}

\begin{figure}
\caption{Comparison of bifurcation diagrams of frequencies $\omega$ (a,d), 
         variance $M$ of the concentration field (b,e), and Nusselt number
         $N$ (c,f) calculated by our Galerkin method and a finite difference
         MAC scheme. Results of the Galerkin method are shown by solid (SOC)
         and dashed (TW) lines. MAC results with a spatial resolution
         of $dx=dz=0.05$ [16]
         are shown by filled circles (TW), open squares (SOC), and open
         triangles (phase fixed, unstable SOCs). More accurate MAC results
         ($dx=dz=0.025$) are displayed as open lozenges. The dotted lines
         show results of a pure fluid,
         $\psi=0$: thick dots -- Galerkin, thin dots -- MAC.}
\label{fig:BifVgl}
\end{figure}

\begin{figure}
\caption{Bifurcation diagrams of $w_{\rm max}^2$ versus
         reduced Rayleigh number $r$ at weak Soret couplings and $L=0.01$,
         $\sigma=10$. Open lozenges represent SOCs, filled lozenges TWs.
         Results were computed with our Galerkin method.}
\label{fig:BifTop}
\end{figure}

\begin{figure}
\caption{Bifurcation diagrams of convective amplitude $w_{\rm max}$ (a) and
         frequency $\omega$ (b) for fluids with $L=0.03$, $\sigma=1$, and
         $\psi=-0.055$. Graph (c) shows eigenvalues determining the
         stability of the TWs: open (filled) symbols correspond to unstable, 
         $\mbox{Re}\gamma_{\rm TW}>0$, (stable, $\mbox{Re}\gamma_{\rm TW}<0$)
         states (stationary: $\mbox{Im}\gamma_{\rm TW}=0$,
         oscillatory: $\mbox{Im}\gamma_{\rm TW}\neq0$). The states
         corresponding to the small filled circles are not shown in (a) and
         (b).}
\label{fig:MTW}
\end{figure}

\begin{figure}
\caption{Phase diagram of the $\psi$--dependence in mixtures with $\L=0.01$
         and $\sigma=10$ in a double logarithmical plot $r-1$ {\em vs.\/}~$\psi$.
         SOC properties are shown by dotted lines (saddle node $r_{\rm
         SOC}^s$: open triangles up, bifurcation $r_{\rm stat}$: open
         triangles down). Solid lines correspond to points in the TW
         bifurcation diagrams (Hopf bifurcation $r_{\rm osc}$: filled
         squares, SOC--TW--transition $r^*$: open lozenges, saddle node
         $r_{\rm TW}^s$: filled circles).}
\label{fig:PhaseGanz}
\end{figure}

\begin{figure}
\caption{Bifurcation diagram of convective amplitude $w_{\rm max}$ and
         phase velocity $v$ {\em vs.~\/}reduced Rayleigh number $r$ for TWs
         in a mixture with $L=0.01$, $\sigma=10$, $\psi=-0.25$, and $k=\pi$.
         The stable (unstable) branches are shown by solid (dotted for
         $w_{\rm max}$ and dashed for $v$) lines. Letters A -- D identify
         states discussed in the text.}
\label{fig:GeschwBif}
\end{figure}

\begin{figure}
\caption{Lateral and vertical concentration profiles of the four states
         identified in Figs.~\ref{fig:BifOverview} and \ref{fig:GeschwBif}.}
\label{fig:Profile}
\end{figure}

\begin{figure}
\caption{Concentration $C$ {\em vs.~\/}streamfunction $\tilde{\phi}$
         in the comoving frame for the states identified
         in Figs.~\ref{fig:BifOverview} and \ref{fig:GeschwBif}. The dotted
         lines referring to the conductive state are explained in the text.
         Since $C$ and $\tilde{\phi}$ was evaluated on a grid some of the
         fine structure seen in the plots reflects the grid discretization.} 
\label{fig:CvonPhi}
\end{figure}

\begin{figure}
\caption{Concentration currents in a TW propagating to the right
         in the comoving frame
         ($L=0.01$, $\sigma=10$, $\psi=-0.25$, and $\omega=4$) with complete
         Soret coupling (left column) and averaged, boundary Soret forcing
         (right column). For details see text. The top row (a) shows the
         total concentration current
         $\tilde{\vec{J}}$
         (\ref{eq:conccurrent}) in the left column and the modified current 
         $\tilde{\vec{J}}_{\rm modified}$ (\ref {eq:modcurrent}) in the right 
         column each
         together with the streamlines of $\tilde{\vec{u}}$.
         The scaling factor for the lengths of the arrows is $0.5$ relative
         to the units of lateral and vertical axes.
         Row (b) shows the influence of the
         Soret coupling more explicitly. Left column: $L\psi\Nabla T$ 
         (complete coupling); right column: 
         $-L\psi N \vec{e}_z$ (averaged coupling). In row (b), the arrow lengths
         had to be magnified by a factor $30$ relative to those of the top
         row in order to make them visible.}
\label{fig:Stroeme}
\end{figure}

\begin{figure}
\caption{Bifurcation diagrams of TW frequencies. Symbols
         refer to correct Soret coupling and full mode field
         representation. Lines were obtained by the laterally averaged
         Soret effect (\ref{eq:modcurrent}) and the reduced temperature
         representation (\ref{eq:Tansatzred}). The unstable states for the
         complete coupling have been dropped for reasons of clarity.
         Parameters are $L=0.01$ and $\sigma=10$.}
\label{fig:RandBulk}
\end{figure}

\end{document}